\documentclass[preprint,12pt]{aastex}
\usepackage{natbib}
\usepackage{epsfig,amsmath,amssymb}
\begin{document}

\title{Stochastic Gravitational Wave Background from Coalescing Binary Black Holes}

\author{Xing-Jiang Zhu$^{1,2}$, E. Howell$^{2}$, T. Regimbau$^{3}$, D. Blair$^{2}$ and Zong-Hong Zhu$^{1}$}

\affil{$^{1}$Department of Astronomy, Beijing Normal University,
Beijing 100875, China; zhuzh@bnu.edu.cn\\
$^{2}$School of Physics, University of Western Australia, Crawley WA
6009, Australia\\
$^{3}$UMR ARTEMIS, CNRS, University of Nice Sophia-Antipolis,
Observatoire de la C\^{o}te d'Azur, BP 4229, 06304, Nice Cedex 4,
France}

\begin{abstract}
We estimate the stochastic gravitational wave (GW) background signal
from the field population of coalescing binary stellar mass black
holes (BHs) throughout the Universe. This study is motivated by
recent observations of BH-Wolf-Rayet star systems and by new
estimates in the metallicity abundances of star forming galaxies
that imply BH-BH systems are more common than previously assumed.
Using recent analytical results of the inspiral-merger-ringdown
waveforms for coalescing binary BH systems, we estimate the
resulting stochastic GW background signal. Assuming average
quantities for the single source energy emissions, we explore the
parameter space of chirp mass and local rate density required for
detection by advanced and third generation interferometric GW
detectors. For an average chirp mass of 8.7$M_{\odot}$, we find that
detection through 3 years of cross-correlation by two advanced
detectors will require a rate density, $r_0 \geq 0.5\hspace{1mm}
\rm{Mpc}^{-3} \rm{Myr}^{-1}$. Combining data from multiple pairs of
detectors can reduce this limit by up to $40\%$. Investigating the
full parameter space we find that detection could be achieved at
rates $r_0 \sim 0.1\hspace{1mm} \rm{Mpc}^{-3} \rm{Myr}^{-1}$ for
populations of coalescing binary BH systems with average chirp
masses of $\sim 15M_{\odot}$ which are predicted by recent studies
of BH-Wolf-Rayet star systems. While this scenario
is at the high end of theoretical estimates, cross-correlation of
data by two Einstein Telescopes could detect this signal under the
condition $r_0 \geq 10^{-3} \hspace{1mm} \rm{Mpc}^{-3}
\rm{Myr}^{-1}$. Such a signal could potentially mask a primordial GW
background signal of dimensionless energy density,
$\Omega_{\rm{GW}}\sim 10^{-10}$, around the (1--500) Hz frequency
range. \vspace{-2mm}
\end{abstract}


\section{Introduction}
Coalescing systems of stellar mass binary black holes (BBHs) are
among the most likely candidates for the first detection of
gravitational waves (GWs) \citep{bbh1,buo03}. Their enormous
predicted luminosities $\sim 10^{23}L_{\odot}$, would allow future
ground-based interferometric GW detectors, such as Advanced LIGO
\citep{aligo} and Advanced Virgo \citep{avirgo} or third generation
instruments such as the Einstein Telescope \citep[ET;][]{et}, to
probe these sources out to Gpc distances. In this paper, we are
motivated by recent increased rate estimates \citep{metal} to
explore the possibility that a population of BBHs could form a
detectable stochastic GW background (SGWB) signal for these
instruments.

SGWBs can result from the superposition of populations of unresolved
primordial \citep{Grishchuk:1974ny} or astrophysical sources
\citep[see][for a recent review]{regimbau11}. Astrophysical SGWB
signals are important for at least two reasons. Firstly, they
contain rich information on the global properties of source
populations, such as their source rate evolution, their mass ranges
and their average energy emissions. Secondly, a dominant continuous
astrophysical background could mask the relic SGWB signal from the
very early Universe \citep[see][for reviews]{Maggiore,Buonanno}.

Mergers of binary neutron stars have been suggested as sources of
potentially detectable SGWBs \citep{Regimbau06,Regimbau07}. Recent
observations of BH-Wolf-Rayet (WR) star systems \citep{Crowther} and
new estimates in the metallicity abundances of star forming galaxies
\citep{2008MNRAS.391.1117P} imply that the Galactic merger rate of
BBHs may be of a similar order to that of binary neutron stars
\citep{metal}. Therefore, a population of more luminous BBHs could
produce a dominant background signal. Our aim is to explore upper
limits for a SGWB from coalescing BBHs over a range of rates and
system masses. We investigate the constraints future ground based
interferometric GW detectors will be able to place on the average
properties of the BBH population.

The paper is organized as follows. In Section 2 we discuss rate
estimates of coalescing BBHs and then derive cosmic source rate
evolution models for different star formation histories and minimal
delay times. In Section 3, source energy spectra for coalescing BBHs
are obtained using the template gravitational waveforms of
\citet{IMR} and \citet{spin_IMR}. We then calculate the BBH
background in Section 4 and discuss the detection regimes,
detectability and constraints on the parameter space of the
predicted background in Section 5. Finally, in Section 6 we present
our conclusions.

\section{Rates of BBH coalescences}
The discovery of BH-WR star systems within close proximity has
increased rate estimates of coalescing BBHs. There are presently two
known systems: NGC300 X-1, which lies at a distance of 1.8 Mpc and
is composed of a $\sim 20 M_{\odot}$ BH and a WR star of $\sim 26
M_{\odot}$ \citep{Crowther}; IC10 X-1 contains a BH of a mass at
least $23 M_{\odot}$ and a $\sim 35 M_{\odot}$ WR star, and lies
within 700 kpc \citep{Prestwich}. As WR stars are the progenitors of
Type Ib/c supernovae, if such systems survive the supernova
explosion, BBH systems will form and eventually coalesce within a
timescale of Gyrs \citep{Bulik08}.

Recent results from Sloan Digital Sky Survey, have indicated that
half of recent star formation involved galaxies with low metallicity
\citep{2008MNRAS.391.1117P}. This has a profound effect on the
coalescence rates of compact binaries containing BHs when one
considers that NGC300 X-1 and IC10 X-1 were both formed in low
metallicity environments.

Survival of BBH systems is highly dependent on whether they can
overcome two key obstacles in their evolution. Firstly, post natal
supernova kicks can disrupt a significant proportion of systems.
Secondly, orbital shrinkage during the common envelope phase when
the larger star transfers mass to its smaller companion, can cause
the stars to merge before they become compact objects. \citet{metal}
have shown through population-synthesis modeling that a lower
metallicity environment can suppress these two effects. Firstly,
observational evidence suggests that larger BH masses, which are
produced at low metallicity, are born with lower kick velocities
\citep{Mirabel,bhmass}. Secondly, in lower metallicity environments,
slower radial expansion occurs during the common envelope phase,
thus increasing binary retention. The greater fraction of systems
that can survive, in combination with a greater detection range from
more massive and hence luminous systems, has increased the detection
prospects of BBHs for ground-based interferometric GW detectors
\citep{metal}.

Previous estimates of the coalescence rate of BBHs formed through
isolated binary evolution in the field have ranged over orders of
magnitude, from $10^{-4}$ to 0.3 Mpc$^{-3}\rm{Myr}^{-1}$ with a
realistic value of $5 \times 10^{-3}\, \rm{Mpc}^{-3}\rm{Myr}^{-1}$
\citep{DCO,lsc_rate}. The effect of metallicity discussed above
increases the realistic estimate to $3.1 \times
10^{-2}\,\rm{Mpc}^{-3}\rm{Myr}^{-1}$ assuming a 50-50 mixture of
solar and $10\%$ solar metallicity and the most stringent
evolutionary scenario with respect to system survival \citep{metal}.
Unless we state otherwise, this rate (denoted as $r_{1}$) is adopted in our
calculations.

A higher estimate of $0.43\,\rm{Mpc}^{-3}\rm{Myr}^{-1}$ (denoted as
$r_{2}$) was obtained by assuming that all systems survive early
merger during the common envelope stage. We note that $r_{2}$ leads
to a detection rate for initial LIGO/Virgo of around 5 events per
year. A recent population synthesis study by \citet{Bulik08},
however, has demonstrated the formation of BBHs with high chirp
masses ($\sim 15 M_{\odot}$) from the two BH-WR systems, and has
yielded a similarly high rate of $0.36 \hspace{0.5mm}\rm{Mpc}^{-3}
\hspace{0.5mm} \rm{Myr}^{-1}$ corresponding to 3.6 detections a
year. They suggest that either currently employed searches are
insensitive to higher mass BBH inspirals or that there is an
additional aspect to the evolution of such systems that has not so
far been considered. To take account of this uncertainty, we take
$r_{2}$ as a higher rate.


\subsection{The cosmic rate evolution model}
The cosmic coalescence rate can be extrapolated from the local rate
density $r_0$ by assuming the rate tracks the star formation rate
(SFR). Explicitly, the differential GW event rate in the redshift
shell $z$ to $z + dz$ can be written as
\begin{equation}
dR = r_{0}e(z)\frac{dV}{dz}dz \label{dR},
\end{equation}

\noindent with $dV$ the cosmology dependent co-moving volume element given by

\begin{equation}
\frac{dV}{dz}=4\pi c\frac{r(z)^{2}}{H(z)},
\end{equation}

\noindent where the Hubble parameter $H(z)=
H_{0}[\Omega_{\Lambda}+\Omega_{m}(1+z)^{3}]^{1/2}$ and $r(z)$ is the
comoving distance related to the luminosity distance by
$d_{L}=r(1+z)$. We use the parameters $H_{0}=100 h\cdot
\rm{km}\hspace{0.5mm} \rm{s}^{-1}\hspace{0.5mm} \rm{Mpc}^{-1}$ with
$h=0.7$, $\Omega_{m}=0.3$ and $\Omega_{\Lambda}=0.7$
\citep{cosmology}.

Source rate density evolution is accounted for by the dimensionless
evolution factor $e(z)$, normalized to unity in our local intergalactic neighbourhood.
Following \citet{Regimbau09}, we define $e(z)=
\dot{\rho}_{\ast,c}(z) / \dot{\rho}_{\ast,c}(0)$ normalized to unity
at $z=0$ where
\begin{equation}
\dot{\rho}_{\ast,c}(z)= \int
\frac{\dot{\rho}_{\ast}(z_{f})}{(1+z_{f})} P(t_{d}) dt_{d},
\end{equation}

\noindent relates the SFR to the BBH coalescence rate. Here,
$\dot{\rho}_{\ast}$ is the SFR density in $M_{\odot}
\hspace{0.5mm}\rm{yr}^{-1} \hspace{0.5mm}\rm{Mpc}^{-3}$, based on
the parametric form of \citet{SFR}, derived from recent measurements
of the galaxy luminosity function. To allow for uncertainties in the
SFR, we also consider a recent model described in \citet{Wilkins},
obtained through measurements of the stellar mass density -- this
model gives a much lower rate for $z
> 1$. The factors $z$ and $z_f$ represent the redshift values of
BBHs merger and BBH system formation
respectively. The delay time for BBHs, $t_d$, is
given by the difference in lookback times between $z_f$ and $z$

\begin{equation}
t_{d}= \int_{z}^{z_f} \frac{dz'}{(1+z')H(z')}.
\end{equation}

\noindent Here $P(t_{d})$ is the probability distribution of delay
times. For this we take the form, $1/t_{d}$, with a lower cutoff
$t_0$ at 100 Myr as suggested by current population synthesis
studies \citep{ps02,ps06,Dominik}. To account for uncertainty in the
lower cutoff and possible correlation between $P(t_{d})$ and BBH
chirp masses, we additionally consider a longer minimal delay time
of  $t_0 = 500$ Myr (T. Bulik, private communication).

Figure 1. shows the cosmic rate evolution factor $e(z)$ of BBH
coalescences for different SFRs and minimal delay times. Although
the models show some variation in peak $z$, we show in section 4
that the magnitude of the background is largely dependent on $r_0$
rather than the form of $e(z)$.

\section{The GW energy spectrum of coalescing BBHs}
The evolution of a coalescing BBH is traditionally divided into
three phases: \emph{inspiral}, \emph{merger} and \emph{ringdown}.
While the early inspiral and ringdown phases can be approximated
analytically by post-Newtonian expansion and perturbation theory, to
model the late inspiral and merger requires a numerical solution of
the Einstein equations. In the last few years breakthroughs in
numerical relativity have enabled the inspiral-merger-ringdown
evolution of BBH coalescences to be modeled with high accuracy for a
broad space of parameters \citep[see, e.g.,][for
reviews]{han09,bbh_status}.

For the complete evolution history of coalescing BBHs,
phenomenological waveforms can be constructed by frequency domain
matching of post-Newtonian inspiral waveforms with coalescence
waveforms from numerical simulations
\citep{ajith07,buo07,pan08,san10}. Such waveforms share a common
feature, in that the Fourier amplitude is approximated to a leading
order as a power-law function of frequency $\nu^{-7/6}$ for the
inspiral phase, followed by $\nu^{-2/3}$ for the merger stage and a
Lorentzian function around the quasi-normal mode ringdown frequency
for the ringdown stage.

For this study, we convert to a energy spectrum to account for the
individual source emissions. We choose two template models: one from
\citet{IMR}, for a non-spinning case, and another from
\citet{spin_IMR} for BBHs with non-precessing spins. In \citet{IMR}
the Fourier amplitude is given by their equation (4.13), which we
convert to an energy spectrum $dE/d\nu$. As we expect the inspiral
spectrum to equal the post-Newtonian approximation \citep[see, e.g.,][]{Cutler93,Finn93}, the inspiral-merger-ringdown spectrum for
a BBH with component masses $m_1$ and $m_2$ is given by

\begin{equation}
\label{dedf}
 \frac{dE}{d\nu} \equiv \frac{(G \pi)^{2/3} M_c^{5/3}}{3}
\left\{\begin{array}{l}
\displaystyle \nu^{-1/3} \hspace{3mm} \mathrm{if}\, \nu < \nu_{1} \\\\
\displaystyle \omega_1 \nu^{2/3} \hspace{3mm} \mathrm{if}\, \nu_{1} \leq \nu < \nu_{2} \\\\
\displaystyle \omega_2 \left[ \frac{\nu}{1+(\frac{\nu-\nu_{2}}{\sigma/2})^2} \right]^{2}\, \hspace{2mm} \mathrm{if}\, \nu_{2} \leq \nu < \nu_{3} \\
\end{array}
 \right.
\end{equation}

Here $M_c$ is the chirp mass, $M_c^{5/3}=m_1 m_2 (m_1+m_2)^{-1/3}$,
$\omega_1=\nu_{1}^{-1}$ and $\omega_2=\nu_{1}^{-1} \nu_{2}^{-4/3}$
are constants chosen to make $dE/d\nu$ continuous across $\nu_{1}$
and $\nu_{2}$. The set of parameters ($\nu_{1}, \nu_{2}, \sigma,
\nu_{3}$) can be determined by the two physical parameters (the
total mass $M$ and the symmetric mass ratio $\eta$) in terms of $(a
\eta^2 + b \eta + c)/\pi M$, with coefficients $a, b, c$ given in
Table 1 of \citet{IMR}, producing (404, 807, 237, 1153)\,Hz for a
$10 M_{\odot}\hspace{-1mm}-\hspace{-1mm}10 M_{\odot}$ BBH ($M_c=8.7
M_{\odot}$).

The waveform presented in \citet{spin_IMR}, includes spin effects
through a single spin parameter $\chi = (1+\delta) \chi_{1}/2 +
(1-\delta) \chi_{2}/2$, with $\delta=(m_1-m_2)/M$ and
$\chi_{i}=S_{i}/m_{i}^2$. The parameter $S_{i}$ represents the spin
angular momentum of the $i$th black hole. The corresponding Fourier
amplitude includes a minor correction (related to $\chi$ and $\eta$)
for non-spinning BBHs. We construct energy spectra for BBHs with
non-precessing spins based on their \mbox{equation (1).}

Figure 2 shows the GW energy spectra for a $10
M_{\odot}\hspace{-1mm}-\hspace{-1mm}10 M_{\odot}$ BBH assuming: the
non-spinning case; $\chi=0.85$; $\chi=0$ and $\chi=-0.85$. The two
extreme values for $\chi$ are set by the numerical simulations of
\citet{spin_IMR}, corresponding to both binary components having
maximal spins aligned or anti-aligned with the orbital angular
momentum. The radiation efficiencies for these energy spectra are
$6.7\%$, $9.74\%$, $5.15\%$ and $4.28\%$ respectively. We note that
the radiated GW energy mainly depends on $M_c$ and $\chi$, and that
the energy spectra for $\nu  \lesssim $ 100 Hz show little
variation. We note that effect of orbital eccentricity is not
considered in our derivation of energy spectrum. This has little
effect on our results as the orbits of coalescing compact objects are expected to circularise \citep{circle} before their GW signals reach the sensitive frequency band of ground-based interferometric detectors \citep{eccen}.

\section{The stochastic GW background}
In this section we evaluate the spectral properties of the BBH
background. It is customary to characterize the SGWB by the energy
density parameter

\begin{equation}
\Omega_{\rm{GW}}(\nu_{\rm{obs}})=\frac{1}{\rho_{c}}\frac{d\rho_{\mathrm{GW}}}{d\ln\nu_{\rm{obs}}},
\end{equation}

\noindent where $\rho_{\rm{GW}}$ is the GW energy density, $\nu_{\rm{obs}}$ is
the observed GW frequency and $\rho_{c}=3H_{0}^{2}/8\pi G$ is the
present value of critical energy density required to close the Universe. For a
SGWB of astrophysical origin, $\Omega_{\rm{GW}}$ is related to the
spectral energy density $F_{\nu}$ (in $\rm{erg}\hspace{0.5mm}
\rm{cm}^{-2}\hspace{0.5mm} \rm{Hz}^{-1}\hspace{0.5mm} s^{-1}$) by

\begin{equation}
\Omega_{\rm{GW}}(\nu_{\rm{obs}})=\frac{\nu_{\rm{obs}}}{c^{3}\rho_{c}}F_{\nu}(\nu_{\rm{obs}})
\label{agwb},
\end{equation}

\noindent where $F_{\nu}$ at the observed frequency $\nu_{\rm{obs}}$ can be
written as

\begin{equation}
F_{\nu}(\nu_{\rm{obs}})=\int_{0}^{z_{\rm{sup}}}
f_{\nu}(\nu_{\rm{obs}},z) \frac{dR}{dz}(z) dz \label{fnu},
\end{equation}

\noindent Here, $z_{\rm{sup}}=
\rm{min}(z_{\rm{max}},\nu_{\rm{max}}/\nu_{\rm{obs}}-1)$, with
$z_{\rm{max}}$ the maximum redshift of SFR model and
$\nu_{\rm{max}}$ the maximal emitting GW frequency. The differential
GW event rate, $dR/dz$, is given by equation (1) and $f_{\nu}$ is
the energy flux per unit frequency emitted by a source at a
luminosity distance $d_{L}(z)$

\begin{equation}
f_{\nu}(\nu_{\rm{obs}},z)=\frac{1}{4 \pi d_{L}(z)^{2}}
\frac{dE}{d\nu}(1+z) \label{flux},
\end{equation}

\noindent where $dE/{d\nu}$ is the gravitational energy spectrum
given by equation (5) and $\nu$ is the frequency in the source frame
which is related to the observed frequency by
$\nu=\nu_{\rm{obs}}(1+z)$.

Figure 3 shows the energy density parameter $\Omega_{\rm{GW}}$ of
the BBH background corresponding to five models outlined in Table 1.
The final model (e) approximates the common power-law behavior of
models (a-d) before reaching a peak of $\sim 10^{-9}$ at 400-600 Hz,
and uses a conservative estimate to account for effects of spin, SFR
and delay time. This model is given by $\Omega_{\rm{GW}}
(\nu_{\rm{obs}}) = 1.95 \times 10^{-11} \nu_{\rm{obs}}^{2/3}$ for 1
Hz$< \nu_{\rm{obs}} <$ 400 Hz. We note that using the two extreme
spectra ($\chi=0.85$ and $\chi=-0.85$) in Figure 2 leads to a
variation in the results of less than $40\%$.

\begin{table}
\begin{center}
\begin{tabular}{llcc}
  Model        & source spectrum     & SFR model & Delay time, $t_0$ (Myr) \\
  \hline
  \hline
  a            & non-spinning    & HB06      & $100$  \\
  b            & $\chi=0$        & HB06      & $100$\\
  c            & non-spinning    & HB06      & $500$ \\
  d            & non-spinning    & W08       & $100$  \\
  e            & power law model &           &  \\
  \hline
  \hline
\end{tabular}
\end{center}
  \vspace{-5mm}
  \caption{The 5 models and their parameters used to determine the energy density parameter $\Omega_{\rm{GW}}$ of a BBH background. Model e, a power law approximation, is described in the text.}
\end{table}


We see in Figure 3 that effects of spin, SFR and delay time are
insignificant for estimating the BBH background considering all
other uncertainties in sources rates and systems masses. Combining
the equations used to calculate $\Omega_{\rm{GW}}$, we find the BBH
background is sensitive to $r_{0}$ and average value of $M_{c}$
through the relation $\Omega_{\rm{GW}} \propto r_0\ M_c ^{5/3}$. It
is useful to express model (e) in the form:

\begin{equation}
\Omega_{\rm{GW}}(\nu_{\rm{obs}}) \simeq 4.2 \times 10^{-10}
\left(\frac{r_0}{3.1 \times
10^{-2}\,\rm{Mpc}^{-3}\rm{Myr}^{-1}}\right)
\left(\frac{M_c}{8.7\hspace{0.5mm} M_{\odot}}\right) ^{5/3}
\left(\nu_{\rm{obs}} \over {100\hspace{0.5mm} \rm{Hz}}\right) ^{2/3}
\label{gwb-bbh}\,,
\end{equation}

\noindent allowing us investigate the parameter space $(r_0, M_c)$
for a SGWB in the next section. We note that equation (10) is
consistent with equation (10) of \citet{GWB_P}. Considering that the
most sensitive frequency regime of the second and third generation
detectors is within 100--200\,Hz, we further adopt a cutoff
frequency of 400 Hz when applying this function.

\section{Issues of detectability}
A number of laser interferometric GW detectors have reached their
design sensitivities and have been coordinating as a global array.
These include the LIGO\footnote{\url{http://www.ligo.caltech.edu/}}
detectors based at Hanford (H) and Livingston (L) in USA, the
Virgo\footnote{\url{http://www.virgo.infn.it/}} (V) detector in
Italy and the
GEO600\footnote{\url{http://www.geo600.uni-hannover.de}} detector in
Germany. The LIGO and Virgo detectors are undergoing upgrades that
will produce a order of magnitude improvement in sensitivity.
Advanced LIGO is expected to be operational by 2015 and an Advanced
Virgo facility is expected to begin commissioning in 2011. Other
plans for advanced detectors include
LCGT\footnote{\url{http://gw.icrr.u-tokyo.ac.jp/lcgt/}} (the
Large-scale Cryogenic Gravitational-wave Telescope; denoted as C),
an underground detector to be built by the TAMA group in Japan and
LIGO-Australia\footnote{\url{http://www.aigo.org.au/}} (A), a 4km
interferometric detector that would be based in Western Australia.
Third generation detectors have also been proposed, such as ET, with
a target sensitivity 100 times greater than current instruments.

To evaluate the detectability of the BBH background, we consider the
five advanced detectors introduced above (H,L,V,C,A). We take
LIGO-Australia to have the same sensitivity as Advanced LIGO and for
ET adopt ET-B sensitivity from \citet{et-B} as well as the latest
ET-D sensitivity from \citet{et-D}. Design sensitivity
curves\footnote{Data are taken from
http://wwwcascina.virgo.infn.it/advirgo/,
https://dcc.ligo.org/cgi-bin/DocDB/ShowDocument?docid=2974,
http://www.et-gw.eu/etsensitivities and \citet{lcgt}.} for these
detectors are shown in Figure 4.

\subsection{Duty Cycle}
For a SGWB of astrophysical origin, in addition to the energy
density parameter and characteristic frequency, another useful
quantity is the duty cycle ($DC$). This is defined as the ratio of
the typical duration of a single signal to the average time interval
between successive events, e.g.,
\begin{equation}
DC=\int_{0}^{z_{\rm{max}}} \bar{\tau}(z)dR(z) \label{duty},
\end{equation}

\noindent where $dR(z)$ is the differential event rate given by
equation (\ref{dR}) and $z_{\rm{max}}$ corresponds to the redshift
limit of considered SFR models. $\bar{\tau}(z)$ is the average
observed duration of GW signals generated by individual sources at
redshift \emph{z}, given at leading order by

\begin{equation}
\bar{\tau}(z)= \frac{5c^5}{256 \pi^{8/3} G^{5/3}} [(1+z)M_c]^{-5/3}
\nu_{\rm{min}}^{-8/3} \label{duration}.
\end{equation}

\noindent Here, $\nu_{\rm{min}}$ is the lower frequency bound of the
detector, which is set by the low-frequency seismic ``wall" for
ground-based interferometric GW detectors. The present LIGO detector
has $\nu_{\rm{min}} \simeq 40$ Hz, and this can be reduced to 10 Hz
for advanced detectors and 1 Hz for ET \citep[see][for
details]{Regimbau09}.

In general, a GW background with a $DC$ of unity or above is defined
as continuous\footnote{We note that to allow for a small number of
events that may be resolved at $DC\sim 1$, some authors consider a
more conservative threshold of $DC\sim 10$ to indicate the
continuous regime \citep{Regimbau09,Eric2010}.}. And non-continuous
signals can be further categorized into popcorn noise ($0.1\leq DC <
1$) and shot noise ($DC < 0.1$) type. As source rate evolution will
increase out to large cosmological volumes, some studies investigate
how \emph{DC} too increases with $z$
\citep{coward_regimbau_06,Eric2010}. In this study we are concerned
with the value of $DC$ as seen at the detector and thus equation
(\ref{duty}) is interpreted as a total value.

For a background signal produced by a source population with an
average chirp mass $\langle M_{c} \rangle = 8.7 \hspace{0.5mm}
M_{\odot}$, for rates $r_{1}$ and $r_{2}$ we find $DC$ values of
0.01 and 0.2 for advanced detectors. Therefore the signal will most
likely be of the shot noise category (or at most popcorn noise for
the higher rate $r_{2}$). For ET type detectors the signal will be
continuous with $DC$ values of 5.8 and 80.6 for rates $r_{1}$ and
$r_{2}$ respectively.

Although for advanced detectors, the SGWB calculated in this paper
is not continuous (Gaussian) even at the higher rate $r_2$, we note
that \citet{drasco} have found the cross correlation method nearly
optimal for a $DC > 10^{-3}$ over 1-year integration\footnote{We
note however, \citet{drasco} also proposed a new statistic to search
for SGWB with low duty cycles, which is currently being investigated
in the LIGO/Virgo collaboration.}. In the following sections we
therefore consider the cross correlation statistic to assess the
detectability of the estimated BBH background.

\subsection{Cross correlation of detectors}
The optimum detection strategy for continuous GW background signals
is cross-correlating the output of two neighbouring detectors
\citep[see,][]{SNR,Maggiore}. This requires that the detectors are
separated by less than one reduced wavelength, which is about 100 km
for frequencies around 500 Hz where $\Omega _{\mathrm{GW}}(f)$ might
peak. The detectors also need to be sufficiently well separated that
their noise sources are largely uncorrelated. We note that although
this may not be possible for ET, techniques are in development to
remove environmental noise and instrumental correlations
\citep{Fotopoulos:2008yq}.

Under these conditions, assuming Gaussian noise in each detector and
optimal filtering, a filter function chosen to maximize the
signal-to-noise ratio, SNR for two such detectors is given by
\citep[][equation 3.75]{SNR}
\begin{equation}
{\mathrm{SNR}^2} =  {9 H_0^4 \over 50 \pi^4} T \int_0^\infty df \>
{\gamma^2 (f) \Omega_{\rm{GW}}^2(f) \over f^6 P_1(f) P_2(f)}\
\label{SNR},
\end{equation}
Here $\gamma (f)$ is the `overlap reduction function', which
accounts for the separation and relative orientation of the
detectors \citep{gammaf}, and $P_1(f)$ and $P_2(f)$ are the noise
power spectral densities of the detectors, and $T$ is the
integration time. As the optimal filter depends on
$\Omega_{\rm{GW}}(f)$, a range of filter functions based on
theoretical expectations of this function will need to be used.

In this study we use data of relative positions and orientations for
10 independent pairs of the five advanced detectors given in Table 3
of \citet{gamma2} and employ the tensor-mode functions described in
their equations (33-35). For ET we assume two detectors of
triangular shape ($60^{\circ}$ between the two arms) and separated
by an angle of $120^{\circ}$, for which the $\gamma(f)$ has a
constant value of $-3/8$ from 1 Hz to 1000 Hz \citep{Eric2010}. We
adopt a value of SNR = 3 to indicate detection, corresponding with
false alarm rate of 10\% and detection rate of 90\% \citep{SNR}. We
also assume an integration time of 3 years for advanced detectors
and 1 year for ET. \vspace{-2mm}

\subsection{Assessing the detectability using the worldwide network}

Calculating the SNRs for SGWB model (e) shown in Figure 3, we find a
value of 0.14 through cross-correlation by two the Advanced LIGO
detectors H-L. For the other four models shown in Figure 3, we find
variation in SNR of within $20\%$. For ET we find SNRs of 59 and 112
assuming ET-B and ET-D sensitivities respectively for model (e),
indicating that this signal will be easily-detected by third
generation detectors. These results, based on average quantities,
suggest that to detect the BBH background with two Advanced LIGO
detectors will require a rate greater than even the higher rate
estimate, $r_{2}\,\sim 0.43\,\rm{Mpc}^{-3}\rm{Myr}^{-1}$. As there
will exist variation in the sensitivities, locations and
orientations of detectors within a worldwide detector network, it is
useful to compare the performances of different detector pairs and
investigate how combining the network could improve the detection
prospects.

Two approaches of combining 2N detectors to increase the sensitivity
of a stochastic background search have been proposed by \citet{SNR}.
We apply these two methods to a network of 4 second generation
detectors. In each case, the optimal SNR can be expressed as
follows, with individual detectors (1-4) indicated in parenthesis:

(i) \emph{Four-detector correlation} (FC) - can be performed by
directly correlating the outputs of 4 detectors

\begin{equation}
\rm{SNR}^2_{\rm{optI}}\approx {}^{(12)} \rm{SNR}^2\ {}^{(34)}
\rm{SNR}^2 + \ {}^{(13)} \rm{SNR}^2\ {}^{(24)} \rm{SNR}^2 + \
{}^{(14)} \rm{SNR}^2\ {}^{(23)} \rm{SNR}^2\,.
\end{equation}

(ii) \emph{Combining multiple pairs} (CP) - is performed by
correlating the outputs of a pair of detectors, and then combining
measurements from multiple detector pairs
\begin{equation}
\rm{SNR}^2_{\rm{optII}} = {}^{(12)} \rm{SNR}^2 + \ {}^{(13)}
\rm{SNR}^2 + \cdots + \ {}^{(34)} \rm{SNR}^2.
\end{equation}

\noindent We now investigate the detectability of the BBH background
by considering both these two approaches, as well as the
cross-correlation method between two detectors. To do this, we
substitute equation (\ref{gwb-bbh}) for $\Omega_{\rm{GW}}$ into
equation (\ref{SNR}). Although there are uncertainties in $r_0$ and
the true ranges of $M_c$, this approximation allows us to quantify
the advantages of the different approaches outlined above. The goal
here is to provide insight into both the requirements for a
detection and the constraints that can be supported by a
null-detection.

Firstly, we set $M_{\rm{c}} = 8.7 \hspace{0.5mm} M_{\odot}$ and
compare standard cross-correlation measurements between different
pairs of the five advanced detectors (H,L,V,C,A). We also determine
any improvements that can be gained from the two approaches
described above -- (i) FC and (ii) CP. We assess performance through
the values of $r_0$ required to produce a SNR=3.

\begin{table}
\caption {Minimum values of the coalescence rate $r_0$ (in
$\rm{Mpc}^{-3} \hspace{0.5mm} \rm{Myr}^{-1}$) to detect the SGWB
from coalescing BBHs assuming an average chirp mass $M_c = 8.7
\hspace{0.5mm} M_{\odot}$. We explore the result from cross
correlation measurements between different pairs of the worldwide
network of second-generation ground-based detectors (H,L,V,C,A) and
for two approaches of combining 4 detectors (FC and CP), taking a
SNR of 3 indicate detection.}
\begin{center}
\begin{tabular}{lccccc}
\hline \hline
 Pair & A-C & A-H & A-L & A-V & C-H \\
      & 1.88 & 0.90 & 0.85 & 2.31 & 3.11 \\
\hline
 Pair & C-L & C-V & H-L & H-V & L-V \\
      & 10.86 & 1.04 & 0.55 & 1.83 & 1.52 \\
\hline
 Combination & ACHL & ACHV & ACLV & AHLV & CHLV \\
     FC     & 1.25 & 1.16 & 1.10 & 1.04 & 0.75 \\
     CP     & 0.40 & 0.57 & 0.56 & 0.38 & 0.45 \\
\hline \hline \label{table_rates}
\end{tabular}
\end{center}
\end{table}

Table \ref{table_rates} outlines the constraints on $r_0$ we obtain
through these different approaches for second generation detectors.
We see that variation in the function $\gamma(f)$ between different
detector pairs together with different sensitivity levels, produces
a variation in the values of $r_{0}$ required for detection. Among
the 10 independent pairs of the five advanced detectors, H-L
performs best as indicated by the lowest required value of $r_0$
($0.55 \hspace{1mm}\rm{Mpc}^{-3} \rm{Myr}^{-1}$). We note that these
values are above the higher rate $r_{2}$. For cross-correlation
measurements between two ETs, minimum detectable values of $r_0 \sim
1.34 \times 10^{-3} \hspace{1mm} \rm{Mpc}^{-3}\rm{Myr}^{-1}$ and
$r_0 \sim 7 \times 10^{-4} \hspace{1mm} \rm{Mpc}^{-3}\rm{Myr}^{-1}$
are obtained using ET-B and ET-D sensitivities respectively.

The values presented by Four-detector correlation, FC, suggest that
no considerable advantage can be obtained through this approach.
However, we see that by combining multiple pairs, CP, produces the
most consistent improvement, the best of which comes from the AHLV
combination (around a $40\%$ improvement on H-L). Equation (15)
shows that applying the CP method to the 3 identical arms of ET will
reduce the minimum detectable values of $r_0$ by a factor of
$\sqrt{3}$.

\subsection{The $r_{0}$--$M_{c}$ parameter space of a detectable BBH background}

We now adopt the same methods as employed in the last section to
explore the parameter space of $r_{0}$--$M_{c}$ required for a
detectable BBH background. To allow for uncertainties in $M_{c}$, we
consider a large range of (4--20) $M_{\odot}$. This range includes
most values of $M_{c}$ within the low metallicity ($0.1 Z_{\odot}$)
distribution of \citet{metal}. We note however, that for the more
realistic scenario that accounts for early common envelope mergers
as the stars pass through the Hertzsprung gap, the range is more
constrained, with $M_{c}$ around (4--9) $M_{\odot}$.

Figure 5 shows the $r_{0}$--$M_{c}$ space for three different
detection scenarios: 1) cross correlation of H-L -- the best
performing pair of advanced detectors; 2) AHLV (through CP) -- the
optimal combination of 4 advanced detectors; 3) cross correlation of
two ET type detectors using two possible sensitivities (ET-B and
ET-D). Zones encompassed by the two rates, $r_{1}$ and  $r_{2}$, are
shown by the shaded areas.

The figure further confirms that advanced detectors are not likely
to detect the BBH background at the rate $r_{1}$. Detection would
require the high rate estimate, $r_{2}$ and $M_{c} \gtrsim 10
M_{\odot}$. Such a range for $r_{0}$--$M_{c}$ has been predicted by
\citet{Bulik08} through studies of the two BH-WR systems, NGC300 X-1
and IC10 X-1. A null detection would therefore confirm a lack of
understanding in the binary evolution of such systems based on the
single source models employed in this study.

In regards to detection strategies for SGWB signals, Figure 5 shows
some improvement in detectability through the use of CP. As already
noted in Table \ref{table_rates}, AHLV can improve the detectability
by $40\%$ compared with only two Advanced LIGO detectors (H-L).

For ET, for the entire range of $M_{c}$ the signal would be
comfortably detected at the lower rate $r_{1}$. A detectable
continuous background requires a rate of order $10^{-3} \hspace{1mm}
\rm{Mpc}^{-3}\rm{Myr}^{-1}$ or above. In comparison with ET-B, we
find that a larger parameter space can be probed by ET-D due to an
improved sensitivity at lower frequencies ($< 20$ Hz).

\section{Conclusions}
In this paper we have estimated the potential contribution of a
population of coalescing BBHs to a SGWB signal. We are motivated by
recent observations of BH-WR star systems \citep{Crowther} and by
new estimates in the metallicity abundances of star forming galaxies
\citep{2008MNRAS.391.1117P} that have suggested the rate of BBHs in
field populations may be greater than previously expected. We base
the single source emissions on energy spectra calculated from recent
parameterized waveforms of \citet{IMR,spin_IMR}. Then, assuming that
the BBH rate traces the SFR with some delay time, we derive cosmic
source rate evolution models and extrapolate our single-source model
out to high redshifts. Rather than a population synthesis approach,
we assume average properties (e.g., masses and spins) for the BBH
population to determine the characteristics of the SGWB signal and
principal parameters to which the BBH background is sensitive.

Our results show that for $M_{c} \lesssim 10 M_{\odot}$, the
background is not likely to be detected through cross-correlation by
two advanced detectors even at $r_{2} \sim
0.43\,\rm{Mpc}^{-3}\rm{Myr}^{-1}$. Only for greater values of
$M_{c}$, as have been predicted to result from BH-WR systems such as
NGC300 X-1 and IC10 X-1 \citep{Bulik08}, there is scope for
detection.

To further assess the detection prospects for second generation
detectors, we have considered the possibility of combining a
worldwide network of advanced detectors to improve the
cross-correlation statistic, namely the methods FC and CP described
in section 5.3. We find that of these two approaches, CP can produce
an improvement of up to $40\%$ against a standard cross-correlation
between two detectors. For the third generation detector, ET, the
signal is accessible with a SNR of 59 and 112 at the lower rate
estimate $r_{1} \sim 3.1 \times 10^{-2}\,\rm{Mpc}^{-3}\rm{Myr}^{-1}$
using ET-B and ET-D sensitivities respectively. This signal could
mask the primordial background signal at below around
$\Omega_{\rm{GW}}\sim 4 \times 10^{-10}$ at $\sim 100$\,Hz.

We note that the rates used in this study are computed assuming
Milky Way type galaxies and the standard formation channel --
isolated binary evolution. Massive binary formation in early
elliptical galaxies is expected to improve the coalescence rate
\citep{elli,os10}. This is particularly important for BBHs due to
the longer delay time. In addition, dynamical formation scenarios in
dense stellar environments can make a significant contribution to
BBH rates. These other formation channels will not only increase the
event rate of coalescing BBHs, but also add additional uncertainty
to the average component masses of the BBH population. For example,
simulations by \citet{sado} suggest that the average chirp mass in
clusters is $\langle M_{\rm{c}} \rangle \sim 20 M_{\odot}$, much
larger than the same found in the field $\langle M_{\rm{c}} \rangle
\sim 7 M_{\odot}$. This might indicate two similar backgrounds
peaking at quite different frequencies. We have shown a minimum
detectable $r_0$ for ET at around $10^{-3} \hspace{1mm}
\rm{Mpc}^{-3} \rm{Myr}^{-1}$, which may be in the range of rate
predictions from dynamical formation scenarios
\citep{oleary07,sado,miller}. Therefore, detection at the
sensitivity of ET could enable these two potential background
signals to be untangled, thus allowing the average properties of the
different populations to be probed.

Additionally, clues to how these two populations contribute to a
confusion background may be provided by Advanced LIGO/Virgo through
single detections in the shot noise ($DC \ll 0.1$) regime. New data
analysis techniques, such as the probability event horizon method
\citep{Coward05,Howell07} which extracts the temporal signature from
a population of transient sources or the maximum likelihood
statistic \citep{drasco}, could prove valuable in interrogating this
regime.

In view of future searches for a primordial
background signal, particularly for ET, simulations incorporating
both dynamical and isolated binary formation channels could prove
useful. These will be considered in a future study. \vspace{-2mm}

\section*{Acknowledgments}
We thank Cole Miller for useful discussions during the early stages
of this study. We also thank Luciano Rezzolla for a careful reading
of the initial manuscript and for providing important feedback on
the gravitational wave waveform parameters of coalescing BBHs. The
authors are also grateful to Tomasz Bulik for insightful comments
which have led to some valuable amendments and for providing plots
of chirp mass-delay time relation, to Krzysztof Belczynski and
Michal Dominik for providing the distribution of the BBH delay time
from StarTrack simulations, and to Ilya Mandel for useful discussion
about the BBH parameters. We thank the anonymous referee for useful
suggestions which improved the clarity and presentation of our
results. Z.X.-J. acknowledges the Australian Research Council and
the W.A. Government Center of Excellence Programme for support of
his visit at UWA where this work had been done. Z.Z.-H. is supported
by the National Science Foundation of China under the Distinguished
Young Scholar Grant 10825313 and by the Ministry of Science and
Technology national basic science Programme (Project 973) under
grant No. 2007CB815401.

\begin{figure}
\plotone{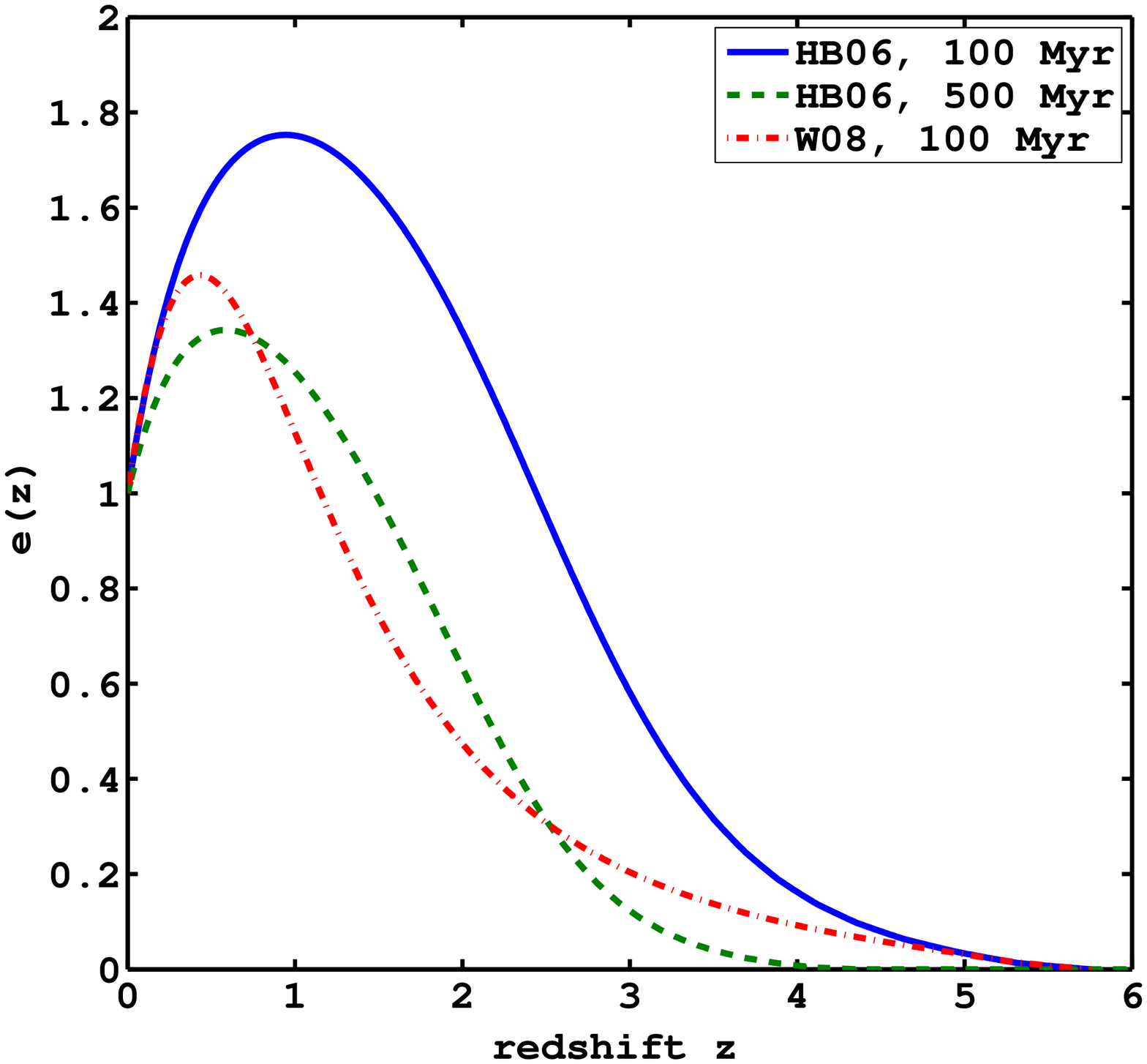} \figcaption{Cosmic rate evolution models
of BBH coalescences. We show the effects of different SFRs by
considering HB06 in \citet{SFR} and W08 in \citet{Wilkins} for a
minimal delay time $t_0= 100$ Myr. To illustrate the effect of $t_0$
we also show the case $t_0 = 500$ Myr for the former SFR.}
\label{rate}
\end{figure}

\begin{figure}
\plotone{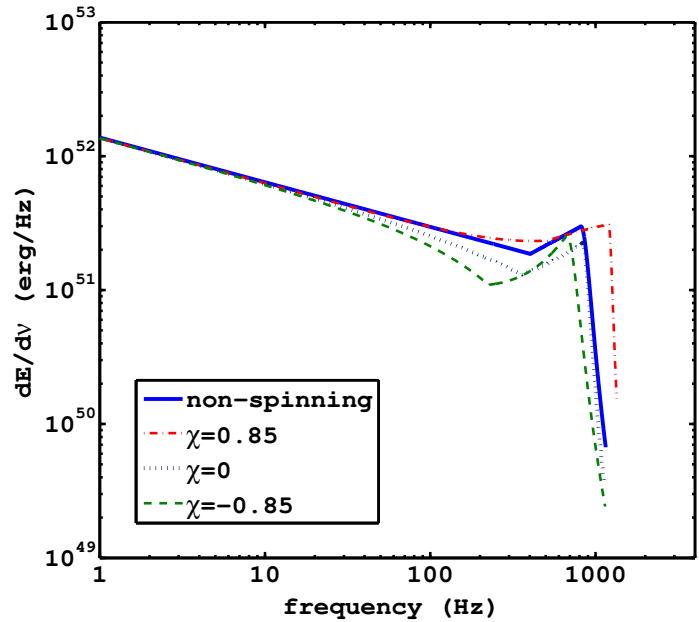} \figcaption{GW energy spectra for a $10
M_{\odot}$--$10M_{\odot}$ coalescing BBH in the non-spinning case
and cases for non-precessing spins with three values of the single
spin parameter $\chi$ (see text).}
\end{figure}

\begin{figure}
\plotone{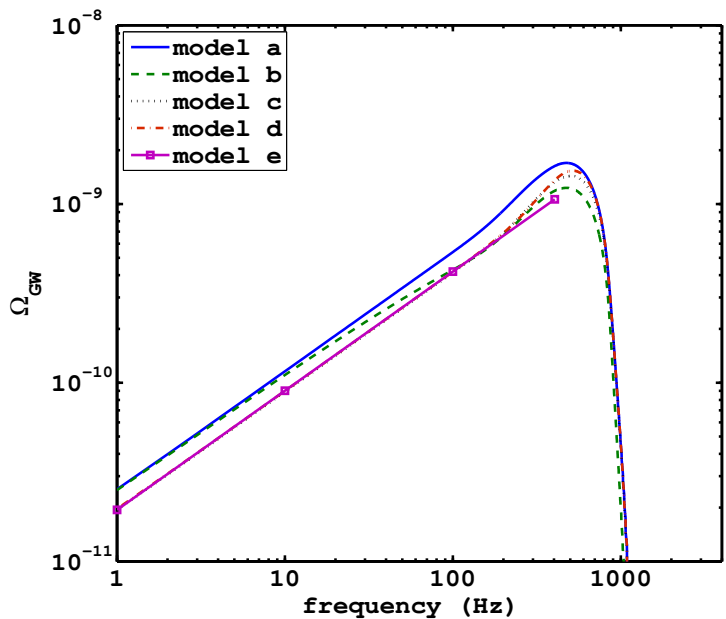} \figcaption{The energy density parameter
$\Omega_{\rm{GW}}$ as a function of observed frequency for the BBH
background corresponding to the five models given in Table 1.}
\end{figure}

\begin{figure}
\plotone{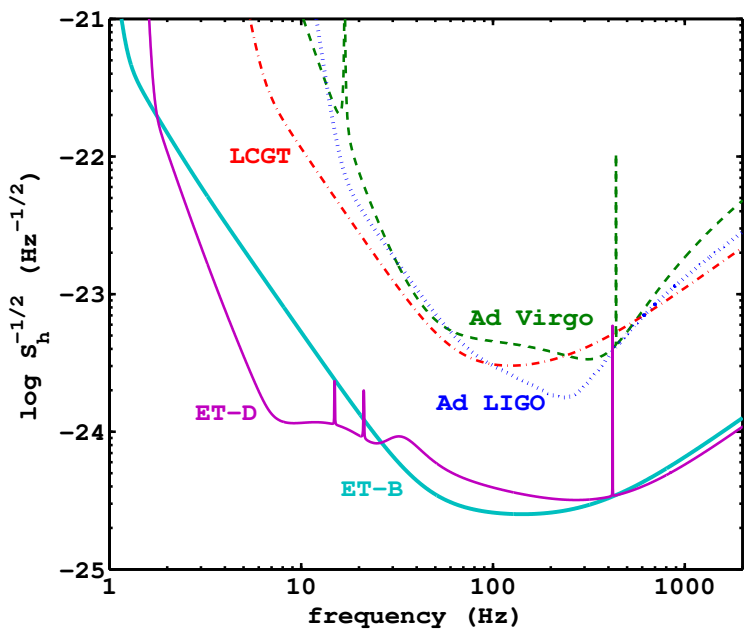} \figcaption{The design sensitivity
curves for future ground-based detectors: Advanced Virgo (Ad Virgo),
LCGT, Advanced LIGO (Ad LIGO) and two possible configurations of the
Einstein Telescope, ET-B and ET-D.} \label{sensitivity}
\end{figure}

\begin{figure}
\plotone{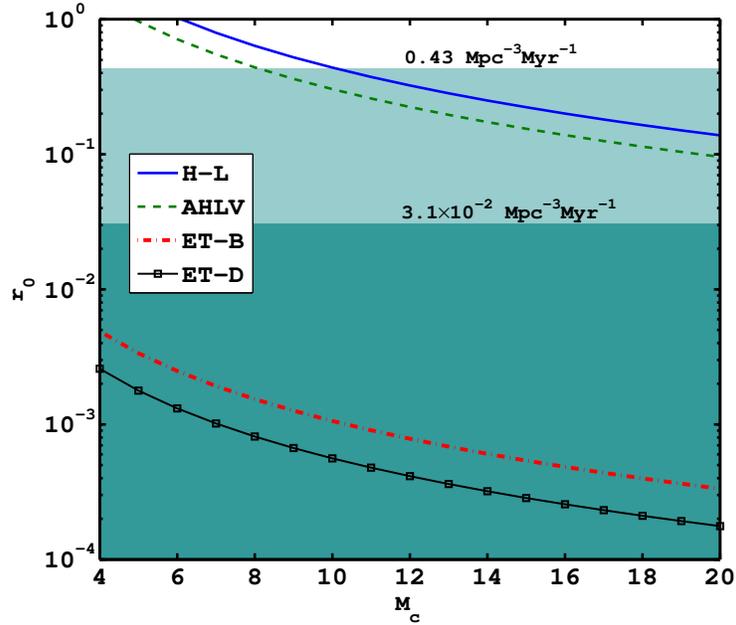} \figcaption{The detectable space of
parameters for a stochastic background formed by coalescing BBHs:
the average chirp mass $M_c$ (in $M_{\odot}$) and the local rate
density $r_0$ (in $\rm{Mpc}^{-3} \hspace{0.5mm} \rm{Myr}^{-1}$). The
$r_{0}$--$M_{c}$ curves correspond with those required to produce a
SNR=3, through cross correlation measurements by two Advanced LIGO
detectors (H-L) and two third generation detectors adopting
ET-B/ET-D sensitivities, and the optimal combination of 4 advanced
detectors (AHLV). The region above these curves can be considered
the detectable parameter space. The two shaded regions correspond to
zones encompassed by the two rates $r_{2}$ and $r_{1}$, based on the
estimates of \citet{metal}.}
\end{figure}

\end{document}